\pdfoutput=1

\documentclass[11pt]{article}

\usepackage[]{ACL2023}
\usepackage{times}
\usepackage[T1]{fontenc}
\usepackage{microtype}
\usepackage{inconsolata}
\usepackage{booktabs}
\usepackage{adjustbox}
\usepackage{multirow}
\usepackage{amsmath}

\title{
Conventional Contrastive Learning Often Falls Short: Improving Dense Retrieval with Cross-Encoder Listwise Distillation and Synthetic Data}

\author{
\bf Manveer Singh Tamber$^{1}$, Suleman Kazi$^{2}$, Vivek Sourabh$^{2}$, Jimmy Lin$^1$
\\[1ex]
$^1$~University of Waterloo \quad $^2$~Vectara
\\[1ex]
        \texttt{\{mtamber, jimmylin\}@uwaterloo.ca}, 
        \texttt{\{suleman, vivek\}@vectara.com}\\
}

\begin{document}
\maketitle

\begin{abstract}

We investigate improving the retrieval effectiveness of embedding models through the lens of corpus-specific fine-tuning.
Prior work has shown that fine-tuning with queries generated using a dataset's retrieval corpus can boost retrieval effectiveness for the dataset.
However, we find that surprisingly, fine-tuning using the conventional InfoNCE contrastive loss often reduces effectiveness in state-of-the-art models.
To overcome this, we revisit cross-encoder listwise distillation and demonstrate that, unlike using contrastive learning alone, listwise distillation can help more consistently improve retrieval effectiveness across multiple datasets.
Additionally, we show that synthesizing more training data using diverse query types (such as claims, keywords, and questions) yields greater effectiveness than using any single query type alone, regardless of the query type used in evaluation.
Our findings further indicate that synthetic queries offer comparable utility to human-written queries for training.
We use our approach to train an embedding model that achieves state-of-the-art effectiveness among BERT embedding models.
We release our model\footnote{\url{https://huggingface.co/manveertamber/cadet-embed-base-v1}} and both query generation and training code to facilitate further research.\footnote{\url{https://github.com/manveertamber/cadet-dense-retrieval}}

\end{abstract}

\section{Introduction}

Fine-tuning dense retrievers remains a challenging problem.
In this work, we focus on enhancing retrieval effectiveness through the lens of fine-tuning state-of-the-art (SOTA) embedding models using synthetic queries tailored to particular corpora.
A dominant training approach for embedding models utilizes the InfoNCE contrastive loss~\cite{oord2018representation}.
However, we show that fine-tuning with this loss often reduces effectiveness in the context of corpus-specific fine-tuning, despite standard refinements such as passage deduplication, hard-negative mining, and negative de-noising.
This highlights limitations in contrastive learning and motivates exploring alternative methods.

To address these limitations, we revisit cross-encoder listwise distillation~\cite{yang2020retriever, rocketqav2}, which we argue is a relatively underused training method in recent work.
Through the lens of corpus-specific fine-tuning, we demonstrate the necessity of cross-encoder listwise distillation, comparing it directly with using contrastive learning alone.
Unlike contrastive learning, which might treat relevance as binary through positive and negative query-passage pairs, listwise distillation leverages scores from a teacher model, offering richer and more nuanced relevance signals.
We demonstrate that the limitations of contrastive learning are not only the result of false negatives.
Even when we filter false negatives using the same cross-encoder teacher, contrastive fine-tuning still generally results in reduced effectiveness.
However, we show that integrating listwise distillation alongside contrastive learning more consistently improves retrieval effectiveness across diverse datasets, even when the cross-encoder teacher has never been trained on the target dataset.

Further, we explore diverse synthetic query generation, generating query types in the form of natural web search queries, questions, titles, claims, and keywords, without relying on query examples from the retrieval datasets.
Our experiments show that training with a diverse mixture of query types outperforms relying on single query types for training across multiple retrieval tasks, regardless of the evaluation query types, by producing more training data.
Notably, we show that these synthetic queries provide comparable utility to human-written queries for training retrieval models.

While recent research has explored LLMs as effective embedding models, we focus on smaller, computationally efficient BERT-base models for their practicality in retrieval systems.
We fine-tune strong BERT-base embedding models including \textit{\small bge-base-en-v1.5}~\cite{bgecpack}, \textit{\small gte-base-en-v1.5}~\cite{gte}, and \textit{\small snowflake-arctic-embed-m-v1.5}~\cite{merrick2024arctic}, which we refer to as BGE, GTE, and Arctic for brevity.
We also analyze the unsupervised variant of E5-base \textit{\small e5-base-unsupervised}~\cite{wang2022text} to investigate fine-tuning a model with strong contrastive pre-training but without further fine-tuning.

Through extensive experiments across diverse embedding models and datasets, we show that our approach consistently enhances retrieval effectiveness.
Leveraging these insights, we train an embedding model that achieves state-of-the-art effectiveness among BERT-based retrievers.
Our work systematically highlights practical and effective strategies for training stronger dense retrieval models, both general-purpose and dataset-specific.

\section{Background}

\subsection{Training Embedding Models}

Leading dense retrieval models across the different model sizes on the MTEB leaderboard~\cite{muennighoff-etal-2023-mteb} for BEIR retrieval~\cite{thakur2021beir} train exclusively with an InfoNCE~\cite{oord2018representation} loss function~\cite{choi2024linq,lee2024nv, bgecpack, gte, merrick2024arctic}.
Typically, while using the InfoNCE loss, in-batch negatives and mined hard-negatives are used to learn to represent text with embeddings contrastively.
Many of the recent works in training dense retrieval models have continued to emphasize the importance of training with hard-negatives to train effective models~\cite{gte,merrick2024arctic,lee2024nv,choi2024linq,nvretriever}.

However, training with hard negatives presents challenges.
Some mined ``negatives'' might be relevant despite lacking labels~\cite{rocketqa}, potentially hindering contrastive learning.
Filtering techniques, such as excluding negatives based on relevance scores from a teacher model~\cite{nvretriever}, can mitigate this issue.
Nevertheless, contrastive loss depends on positive and negative query-passage pairs, simplifying the often non-binary nature of passage relevance.

An alternative training approach is knowledge distillation from cross-encoder teachers~\cite{yang2020retriever, hofstatter2020improving, rocketqav2}.
In particular, listwise distillation, which minimizes the KL divergence between the score distributions from a cross-encoder and a bi-encoder model~\cite{yang2020retriever}, has shown strong effectiveness~\cite{menon2022defense}, outperforming a margin-based loss~\cite{hofstatter2020improving}.

Listwise distillation has been used to train TCT-ColBERT~\cite{lin2020distilling}, with ColBERT as the teacher~\cite{colbert}, and in the E5 models~\cite{wang2022text}, though the specific cross-encoder teacher was not disclosed.

A notable alternative approach involves distilling the embeddings directly from one or multiple teacher embedding models~\cite{tamber2024can}.
This approach has been used to train the Stella~\cite{zhang2024jasper} models.

\subsection{Corpus-Specific Fine-Tuning}

The corpus-specific fine-tuning of retrievers often relies on synthetic query generation~\cite{liang2020embedding, ma2020zero}. The idea is to generate realistic queries for a target corpus and then fine-tune a retriever on these generated synthetic query-passage pairs.

~\citet{thakur2021beir} explored using a T5-based query generator~\cite{t5, docTTTTTquery} trained on MSMARCO~\cite{bajaj2016ms}.
Fine-tuning the TAS-B dense retriever~\cite{tasb} with these queries yielded mixed results, sometimes underperforming the original TAS-B and BM25.
GPL~\cite{gpl} extended this by using the same query generator, but adding a cross-encoder labelling step, training the retriever to mimic the score margins between query-passage pairs assigned by the cross-encoder.

More recent methods leverage LLMs for zero-shot query generation.
InPars~\cite{inpars} used GPT-3~\cite{gpt3} for few-shot query generation to adapt rerankers, later switched to the open-source GPT-J~\cite{gpt-j} in InPars-v2~\cite{inparsv2}.
Promptagator~\cite{Promptagator} used few-shot examples of task-specific queries and passages to generate queries more in line with those from the BEIR datasets being used for evaluation.
UDAPDR~\cite{saad2023udapdr} focused on the corpus-specific fine-tuning of ColBERTv2~\cite{colbertv2}, a multi-vector retrieval model, leveraging GPT-3 and FLAN-T5 XXL~\cite{flant5} for query generation to train up to 10 cross-encoders that were each in turn used to annotate triples of (query, positive document, and negative document) to fine-tune retrieval models.

We employ the lightweight Llama-3.1 8B model~\cite{dubey2024llama} to generate a diverse range of query types (e.g., questions, claims, titles, keywords), without relying on queries from evaluation datasets.
Inspired by work highlighting the benefits of query diversity for generalization~\cite{lin-etal-2023-train}, we show that the diversity in our generated queries promotes generalization across tasks and domains and that these synthetic queries can be as effective for training as human-written ones.

\section{Methodology}

\subsection{Synthetic Query Generation}

We generate synthetic queries using randomly sampled passages from a target corpus, limited to 100k passages for practicality.
Our approach assumes that sampled passages sufficiently represent the retrieval task and what users might search for.
We use Llama-3.1 (8B) for query generation, providing up to three examples of query-passage pairs as prompts.
We experiment with six query types: questions, claims, titles, keywords, zero-shot natural user search queries (no examples provided), and few-shot natural user search queries (examples from the MSMARCO training set).
Otherwise, to generate examples, we take Wikipedia passages from BEIR's Natural Questions Wikipedia corpus and use GPT4o~\cite{achiam2023gpt} to generate high-quality queries to use as static examples to guide the generation of Llama-3.1 (8B).
For all experiments in this work, except those explicitly analyzing query type utility (Section~\ref{sec:eval_query_quality}), we use all six query types during training.

Prior corpus-specific fine-tuning work primarily generated synthetic questions or search queries resembling human-written ones from MSMARCO~\cite{saad2023udapdr, thakur2021beir, gpl, inpars}.
Promptagator~\cite{Promptagator} focused on generating task-specific queries, but did so by using examples from the same retrieval task that the models were evaluated on.
In contrast, our approach produces a diverse set of query types without relying on queries from the target evaluation dataset, enabling more generalizable corpus-specific fine-tuning.
We later evaluate the utility of this diverse generation strategy.

All query-generation prompts are available in our GitHub repository.
In general, we instruct the model to produce queries under 20 words for which the associated passage is relevant.

\subsubsection{Filtering Generated Queries}
\label{sec:filtering}

To ensure high-quality training data and uphold the assumptions of contrastive learning (i.e., that positives are semantically closer to queries than negatives), we apply a two-stage filtering process.

First, we eliminate low-quality queries by discarding those whose passage does not appear in the top 20 retrieved results using the retriever to be fine-tuned.
This also reduces the number of queries passed to the more expensive cross-encoder.
Then, we further filter out queries where the corresponding passage is not ranked first among the top 20 retrieved results by the reranker.

While discarding any queries for which the corresponding passage doesn't rank first may be stricter than necessary, especially for listwise distillation, which doesn’t assume binary relevance labels, we apply it uniformly for a fair comparison between contrastive learning and listwise distillation.

\subsection{Model Training}

\begin{table}[t]
\centering
\adjustbox{max width=\columnwidth}{
\begin{tabular}{l|c|ccc}
\toprule
\textbf{Dataset} & \textbf{BGE-Retriever} & \textbf{RankT5-3B} & \textbf{Gemma} \\

\midrule
DL19            & 70.2 & 75.1 & 76.3 \\
DL20            & 67.7 & 77.2 & 77.5 \\
TREC-COVID      & 78.1 & 85.7 & 81.8 \\
NFCorpus        & 37.3 & 40.4 & 41.0 \\
FiQA            & 40.6 & 52.6 & 58.1 \\
SCIDOCS         & 21.7 & \textbf{19.3} & 21.8 \\
ArguAna         & 63.6 & \textbf{34.9} & 87.8 \\
Touché-2020     & 25.7 & 38.6 & 31.2 \\
DBPedia         & 40.7 & 48.1 & 49.7 \\
FEVER           & 86.3 & \textbf{84.9} & 93.1 \\
Climate-FEVER   & 31.2 & \textbf{27.3} & 45.2 \\
SciFact         & 74.1 & 77.1 & 79.5 \\
\bottomrule
\end{tabular}
}
\caption{\textbf{NDCG@10} Scores for the BGE-base retriever and both the RankT5-3B and the Gemma rerankers ranking the top-100 retrieved passages. Reranker scores that fail to improve upon the retriever scores are bolded.}
\label{tab:rerank_ndcg}
\end{table}

\subsubsection{Cross-Encoder Teacher}

We use two cross-encoder models as teachers: \textit{\small{RankT5-3B}}~\cite{zhuang2022rankt5} and \textit{\small{bge-rerankerv2.5-gemma2-lightweight}}~\cite{limaking}.
For domain-specific adaptation experiments, we rely exclusively on RankT5-3B, which is trained only on MSMARCO and Natural Questions (NQ)~\cite{nq}.
This avoids potential dataset overlaps with evaluation sets. 

\begin{table*}[ht!]
    \centering
    \adjustbox{max width=\textwidth}{
    \begin{tabular}{l|cccc|cc|cc|cc|cc}
        \toprule
         & \multicolumn{4}{c}{MSMARCO} & \multicolumn{2}{c}{SciFact} & \multicolumn{2}{c}{FiQA} & \multicolumn{2}{c}{TREC-COVID} & \multicolumn{2}{c}{NFCorpus} \\
        \cmidrule(lr){2-5} \cmidrule(lr){6-7} \cmidrule(lr){8-9}
        \cmidrule(lr){10-11}
        \cmidrule(lr){12-13}
         & \multicolumn{2}{c}{DL19} & \multicolumn{2}{c}{DL20} &  &  &  &  \\
        \cmidrule(lr){2-3} \cmidrule(lr){4-5}
         & NDCG & Recall & NDCG & Recall & NDCG & Recall & NDCG & Recall & NDCG & Recall & NDCG & Recall \\
        \midrule
        BGE & 70.2 & 60.9 & 67.7 & 71.5 & 74.1 & 96.7 & 40.6 & 74.2 & 78.1 & 14.1 & 37.3 & 33.7 \\
        + FT with contrastive loss & 68.6 & 61.0 & 66.8 & 70.8$\dagger$ & 72.1$\dagger$ & 96.0 & 39.5$\dagger$ & 70.7$\dagger$ & 77.7$\dagger$ & 14.0$\dagger$ & 36.3$\dagger$ & 33.0$\dagger$ \\
        + FT with combined loss & \textbf{71.8} & \textbf{63.0} & \textbf{69.7} & \textbf{75.5} & \textbf{76.2} & \textbf{97.0} & \textbf{44.3} & \textbf{75.8} & \textbf{82.2} & \textbf{15.8} & \textbf{38.3} & \textbf{35.8} \\
        \midrule
        GTE & \textbf{71.9} & 62.1 & 71.5 & 69.8 & 75.5 & 97.3 & 48.7 & \textbf{81.7} & 75.3 & 14.0 & 35.3 & 33.1 \\
        + FT with contrastive loss & 71.5 & 62.6 & 67.8$\dagger$ & 69.8$\dagger$ & 71.2$\dagger$ & 95.5 & 45.1$\dagger$ & 77.1$\dagger$ & \textbf{80.2} & 14.3 & 33.3$\dagger$ & 32.8$\dagger$ \\
        + FT with combined loss & 71.8 & \textbf{65.6} & \textbf{72.6} & \textbf{73.3} & \textbf{76.3} & \textbf{97.7} & \textbf{49.5} & {81.4} & {78.2} & \textbf{14.8} & \textbf{37.9} & \textbf{34.6} \\
        \midrule
        Arctic & \textbf{74.4} & 64.7 & 72.1 & 74.2 & 70.5 & 94.8 & 42.5 & 74.8 & \textbf{82.2} & 14.8 & 36.1 & 32.4 \\
        + FT with contrastive loss & 72.2 & 64.3$\dagger$ & 71.9 & 73.1$\dagger$ & 69.1$\dagger$ & 94.9 & 41.8$\dagger$ & 73.0$\dagger$ & 76.6$\dagger$ & 13.5$\dagger$ & 35.3$\dagger$ & 30.8$\dagger$ \\
        + FT with combined loss & 74.3 & \textbf{67.2} & \textbf{73.6} & \textbf{75.4} & \textbf{73.6} & \textbf{95.7} & \textbf{45.5} & \textbf{75.9} & {80.9} & \textbf{14.9} & \textbf{37.4} & \textbf{33.7} \\
        \midrule
        E5-unsupervised & 56.3 & 52.6 & 54.6 & 62.1 & 74.3 & \textbf{98.7} & 40.1 & 71.8 & 60.6 & 11.9 & 34.6 & 32.3 \\
        + FT with contrastive loss & 67.9 & 61.7 & 68.2$\dagger$ & 71.7$\dagger$ & 72.3$\dagger$ & 95.0$\dagger$ & 40.5$\dagger$ & 73.8$\dagger$ & 75.0$\dagger$ & 13.5$\dagger$ & 36.1$\dagger$ & 32.5$\dagger$ \\
        + FT with combined loss & \textbf{72.4} & \textbf{63.3} & \textbf{73.2} & \textbf{75.2} & \textbf{76.3} & 97.0 & \textbf{45.4} & \textbf{77.0} & \textbf{82.4} & \textbf{15.8} & \textbf{37.5} & \textbf{35.5} \\
        \bottomrule
    \end{tabular}
    }
    \caption{NDCG@10 and Recall@100 for the original and fine-tuned (FT) models, examining the effect of the training loss. A $\dagger$ indicates a significant difference between the fine-tuned models trained with the contrastive loss and the combined loss, based on a one-sided, paired t-test $(p < 0.05)$, with Holm--Bonferroni correction across all datasets and metrics for each model's results. Best scores for each model are bolded.}
    \label{tab:loss_results}
\end{table*}

The Gemma reranker is trained on a broader mix of datasets, including MSMARCO, NQ,  ArguAna~\cite{arguana}, HotpotQA~\cite{yang-etal-2018-hotpotqa}, FEVER~\cite{fever}, as well as other datasets such as PubMedQA~\cite{jin2019pubmedqa}, which might confer an advantage on BEIR datasets such as SciFact and SCIDOCS because all three datasets involve retrieval from PubMed.
Despite its narrower training set, RankT5-3B has shown strong effectiveness on out-of-domain tasks~\cite{zhuang2022rankt5, qin2023large, tamber2023scaling, listt5}, making it a reliable teacher for fine-tuning across domains.
The Gemma reranker is only used for our general-purpose training experiments (Section~\ref{sec:general}).

For each synthetic query, we first retrieve the top 20 passages using the retriever model being fine-tuned (or BGE-base for E5-unsupervised).
These candidate passages are then scored by the cross-encoder, and the query is retained only if its original passage ranks first (as described in Section~\ref{sec:filtering}).
Relevance scores from the cross-encoder are normalized using min-max normalization, clipping at the 1st and 99th percentiles. These normalized scores are used in both listwise distillation and for filtering hard negatives during contrastive training.

\subsubsection{Contrastive Loss}

For each generated query, we retrieve the top 20 passages using the retriever being fine-tuned, yielding up to $K = 19$ hard negatives $p_{jk}$ per query $q_j$.
These hard negatives, along with all in-batch passages, serve as negatives for other queries in the batch. The loss is computed as:

\small
\begin{equation}
-\frac{1}{n}{\sum_{i=1}^n \log \frac{e ^ {\frac{s_{de}(q_i,p_i)}{\tau}}}{\sum_{j=1}^n (e ^ {\frac{s_{de}(q_i,p_j)}{\tau}} + \sum_{k=1}^K e ^ {\frac{s_{de}(q_i,p_{jk})}{\tau}})}}
\end{equation}
\normalsize

To filter hard negatives, we apply threshold-based filtering based on normalized scores from the cross-encoder, similar to prior work~\cite{nvretriever}.
We mask out any hard negative whose score exceeds 60\% of the corresponding positive passage's score.
This threshold was selected based on tuning experiments with E5-unsupervised trained on MSMARCO passages and generated queries, evaluated on DL19 and DL20 (see Figure~\ref{fig:threshold_tuning} in the appendix).
We consistently apply this threshold across all models fine-tuned with contrastive loss. 
For any query, passages filtered out as hard negatives are also prevented from re-appearing as in-batch negatives for that query.

We use a temperature of $\tau = 0.01$, following common practice~\cite{bgecpack, gte, merrick2024arctic, wang2022text}.

\begin{table*}[ht!]
\centering
\adjustbox{max width=\textwidth}{
\begin{tabular}{l|cc|cc|cc|c}
\toprule
\textbf{Dataset} 
& \multicolumn{2}{c|}{\textbf{BGE}} 
& \multicolumn{2}{c|}{\textbf{GTE}} 
& \multicolumn{2}{c|}{\textbf{Arctic}} 
& \textbf{Promptagator} \\
& nDCG & Recall & nDCG & Recall & nDCG & Recall & nDCG \\
\midrule
DL19          
& $70.2 \;\rightarrow\; \mathbf{71.8}$ 
& $60.9 \;\rightarrow\; \mathbf{63.0}$ 
& $\mathbf{71.9} \;\rightarrow\; 71.8$ 
& $62.1 \;\rightarrow\; \mathbf{65.6}$ 
& $\mathbf{74.4} \;\rightarrow\; 74.3$ 
& $64.7 \;\rightarrow\; \mathbf{67.2}$ 
& -- \\
DL20          
& $67.7 \;\rightarrow\; \mathbf{69.7}$ 
& $71.5\dagger \;\rightarrow\; \mathbf{75.6}$ 
& $71.5 \;\rightarrow\; \mathbf{72.6}$ 
& $69.8 \;\rightarrow\; \mathbf{73.3}$ 
& $72.1 \;\rightarrow\; \mathbf{73.6}$ 
& $74.2 \;\rightarrow\; \mathbf{75.4}$ 
& -- \\
TREC-COVID    
& $78.1 \;\rightarrow\; \mathbf{82.2}$ 
& $14.1\dagger \;\rightarrow\; \mathbf{15.8}$
& $75.3 \;\rightarrow\; \mathbf{78.2}$
& $14.0 \;\rightarrow\; \mathbf{14.8}$
& $\mathbf{82.2} \;\rightarrow\; 80.9$
& $14.8 \;\rightarrow\; \mathbf{14.9}$
& 75.6 \\
NFCorpus      
& $37.3 \;\rightarrow\; \mathbf{38.3}$ 
& $33.7\dagger \;\rightarrow\; \mathbf{35.8}$ 
& $35.3\dagger \;\rightarrow\; \mathbf{37.9}$ 
& $33.1 \;\rightarrow\; \mathbf{34.6}$ 
& $36.1 \;\rightarrow\; \mathbf{37.4}$ 
& $32.4\dagger \;\rightarrow\; \mathbf{33.7}$ 
& 33.4 \\
FiQA          
& $40.6\dagger \;\rightarrow\; \mathbf{44.3}$ 
& $74.2 \;\rightarrow\; \mathbf{75.8}$ 
& $48.7 \;\rightarrow\; \mathbf{49.5}$ 
& $\mathbf{81.7} \;\rightarrow\; 81.4$ 
& $42.5\dagger \;\rightarrow\; \mathbf{45.5}$ 
& $74.8 \;\rightarrow\; \mathbf{75.9}$ 
& 46.2 \\
ArguAna       
& $\mathbf{63.6} \;\rightarrow\; 61.4$ 
& $\mathbf{99.2} \;\rightarrow\; \mathbf{99.2}$ 
& $\mathbf{62.1} \;\rightarrow\; 60.9$ 
& $99.2 \;\rightarrow\; \mathbf{99.4}$ 
& $56.5\dagger \;\rightarrow\; \mathbf{58.6}$ 
& $98.4\dagger \;\rightarrow\; \mathbf{99.2}$ 
& 59.4 \\
Touch{\'e}-2020   
& $25.7\dagger \;\rightarrow\; \mathbf{35.3}$ 
& $48.7 \;\rightarrow\; \mathbf{50.6}$
& $27.5 \;\rightarrow\; \mathbf{32.1}$
& $48.3 \;\rightarrow\; \mathbf{49.0}$
& $33.2\dagger \;\rightarrow\; \mathbf{37.9}$
& $50.0 \;\rightarrow\; \mathbf{52.4}$
& 34.5 \\
DBPedia       
& $40.7\dagger \;\rightarrow\; \mathbf{45.5}$ 
& $53.0\dagger \;\rightarrow\; \mathbf{56.3}$ 
& $36.9\dagger \;\rightarrow\; \mathbf{44.7}$ 
& $46.9\dagger \;\rightarrow\; \mathbf{55.2}$ 
& $44.7 \;\rightarrow\; \mathbf{45.1}$ 
& $\mathbf{58.7} \;\rightarrow\; 57.2$ 
& 38.0 \\
SCIDOCS       
& $\mathbf{21.7} \;\rightarrow\; 19.4$ 
& $\mathbf{49.6} \;\rightarrow\; 45.5$ 
& $\mathbf{21.7} \;\rightarrow\; 20.2$ 
& $\mathbf{50.1} \;\rightarrow\; 46.6$ 
& $\mathbf{20.0} \;\rightarrow\; 18.6$ 
& $42.3\dagger \;\rightarrow\; \mathbf{43.5}$
& 18.4 \\
FEVER         
& $\mathbf{86.3} \;\rightarrow\; 80.0$ 
& $\mathbf{97.2} \;\rightarrow\; 95.7$ 
& $\mathbf{92.1} \;\rightarrow\; 82.4$ 
& $\mathbf{97.5} \;\rightarrow\; 96.0$ 
& $\mathbf{85.6} \;\rightarrow\; 80.4$ 
& $\mathbf{97.6} \;\rightarrow\; 96.1$ 
& 77.0 \\
Climate-FEVER 
& $\mathbf{31.2} \;\rightarrow\; 25.1$ 
& $\mathbf{63.6} \;\rightarrow\; 59.3$ 
& $\mathbf{40.1} \;\rightarrow\; 30.3$ 
& $\mathbf{71.7} \;\rightarrow\; 63.8$ 
& $\mathbf{34.7} \;\rightarrow\; 27.2$ 
& $\mathbf{66.7} \;\rightarrow\; 63.1$ 
& 16.8 \\
SciFact       
& $74.1 \;\rightarrow\; \mathbf{76.2}$ 
& $96.7 \;\rightarrow\; \mathbf{97.0}$ 
& $75.5 \;\rightarrow\; \mathbf{76.3}$ 
& $97.3 \;\rightarrow\; \mathbf{97.7}$ 
& $70.5\dagger \;\rightarrow\; \mathbf{73.6}$ 
& $94.8 \;\rightarrow\; \mathbf{95.7}$ 
& 65.0 \\
\bottomrule
\end{tabular}
}
\caption{nDCG@10 and Recall@100 for the models (before fine-tuning 
$\rightarrow$ after fine-tuning).
A $\dagger$ indicates a significant difference between the fine-tuned models and the base models, based on a one-sided, paired t-test $(p < 0.05)$, with Holm--Bonferroni correction across all datasets and metrics for each model's results. Best scores for each model are bolded.
Promptagator scores 
(nDCG only) are shown on the right for reference.}
\label{tab:full_training}

\end{table*}

\subsubsection{Cross-encoder Listwise Distillation}

We follow listwise distillation formulations from previous work~\cite{yang2020retriever,rocketqav2} to distill cross-encoder teachers to retrievers.
Given a relevance distribution:

\small
\begin{equation}
\tilde s(q_i,p_i) = {\frac{e ^ {\frac{s(q_i,p_i)}{\tau}}}{e ^ {\frac{s(q_i,p_i)}{\tau}} + \sum_{k=1}^K e ^ {\frac{s(q_i,p_{ik})}{\tau}}}}
\end{equation}
\normalsize

\noindent based on a scoring function $s_{de}(q,p)$ for the dense retriever, which is the cosine similarity of the query and passage embeddings and $s_{ce}(q,p)$ for the cross-encoder, which is the relevance score from the cross-encoder for a passage with respect to a query.
We minimize the KL divergence of the two distributions $\tilde s_{de}(q,p)$ and $\tilde s_{ce}(q,p)$ over all passages retrieved for each query, which include the positive passage and the $K=19$ hard-negative candidate passages.
While we filter false negatives when training with the contrastive loss as described above, there is no need to filter negatives when training with listwise distillation, since it does not rely on positive and negative query-passage pairs.

\subsubsection{Training and Hyperparameter Selection}
\label{sec:hyperparameters}

All models are trained using a learning rate of $2\text{e}{-4}$ with a batch size of 4096 queries.
To support large contrastive batches, we use GradCache~\cite{gao2021scaling}, following prior work showing that larger batches improve retrieval effectiveness~\cite{gte}.
We train for up to 30 epochs using a 90/10 train/dev split, selecting the checkpoint with the lowest loss on the development set.

Hyperparameters were selected by training the unsupervised E5 on MSMARCO passages and generated queries with distillation from RankT5-3B and evaluation on DL19 and DL20, including temperature parameters of $\tau=0.05$ for $\tilde s_{de}(q,p)$ and $\tau=0.3$ for $\tilde s_{ce}(q,p)$ for distillation.
When training with listwise distillation, we train models using the sum of the contrastive loss and the listwise distillation loss, finding that a combination works well with a $0.1$ weight on the contrastive loss.

We apply the same hyperparameters across all datasets and models (BGE, GTE, Arctic, and E5), ensuring consistency and avoiding overfitting to any specific setting.
While different rerankers may have different scoring distributions, we observe that the selected temperature values generalize well and apply them throughout, including in Section~\ref{sec:general} when we distill using the Gemma-reranker.

\subsection{Evaluation}

We evaluate retrieval effectiveness after fine-tuning retrievers on BEIR~\cite{thakur2021beir} corpora and the MSMARCO~\cite{bajaj2016ms} passage ranking dataset.
Effectiveness is measured using NDCG@10 and Recall@100.

For BEIR, we focus on a diverse subset of datasets: TREC-COVID~\cite{voorhees2021trec}, NFCorpus~\cite{NFCorpus}, FiQA~\cite{fiqa}, SCIDOCS~\cite{scidocs}, ArguAna~\cite{arguana}, Touché-2020~\cite{touche}, DBPedia~\cite{dbpedia}, FEVER~\cite{fever}, CLIMATE-FEVER~\cite{diggelmann2020climate}, and SciFact~\cite{wadden-etal-2020-fact}.
We select these datasets because they are under open licenses and represent a wide range of retrieval scenarios, varying in query type (e.g., factual claims, opinion-based questions), corpus (e.g., Wikipedia, scientific abstracts, forum posts), and topic (e.g., finance, COVID-19, climate change).

To evaluate in-domain effectiveness, we fine-tune on the MSMARCO passage dataset.
For this task, we sample 200K passages (rather than 100K as used elsewhere) to better take advantage of its scale and generality.
Evaluation is conducted using the TREC Deep Learning tracks DL19 and DL20~\cite{craswell2020overview, craswell2021overview}.

\begin{table*}[t]
    \centering
    \adjustbox{max width=0.8\textwidth}{
\begin{tabular}{l | c c c c c c}
        \toprule
        \textbf{Synthetic Query Type} 
            & \multicolumn{6}{c}{\textbf{NDCG@10}} \\
        \cmidrule(lr){2-7}
         & DL19 & DL20 & FiQA & SciFact & TREC-COVID & NFCorpus \\

        \midrule
        (No fine-tuning)
            & 56.3 & 54.6 & 40.1 & 74.3 & 60.6 & 34.6 \\
        \midrule
        User Queries Zero-Shot
            & \textbf{\underline{73.0}} & \underline{72.5} & 42.6 & \underline{75.5} & \underline{81.7} & 36.7 \\
        User Queries Few-Shot
            & 72.7 & \underline{72.5} & \underline{43.0} & 74.4 & 80.4 & 36.2 \\
        Questions 
            & 71.5 & 69.8 & 42.3 & 72.9 & 77.5 & 35.5 \\
        Claims 
            & 65.7 & 68.1 & 40.1 & 74.7 & 79.3 & 34.3 \\
        Titles 
            & 72.3 & 69.8 & 42.5 & 72.8 & 79.6 & \underline{37.0} \\
        Keywords 
            & 67.5 & 64.1 & 37.4 & 71.4 & 79.7 & 35.9 \\
        All Query Types
            & 72.6 & 72.2 & 41.9 & 73.5 & 81.0 & 36.5 \\
        \midrule
        All Query Types (No Downsampling)
            & 72.4 & \textbf{73.2} & \textbf{45.4} & \textbf{76.3} & \textbf{82.4} & \textbf{37.5} \\
        \bottomrule
    \end{tabular}
    }
    \caption{Retrieval effectiveness when training the E5-unsupervised model with the different query types considered and all generated queries. Best scores overall are bolded and we underline the best scores when training with a single query type.}
    \label{tab:query_type}

\end{table*}

\subsubsection{Prior Fine-Tuning}
\label{sec:prior_tuning}

Table~\ref{tab:rerank_ndcg} shows NDCG@10 after reranking passages retrieved by BGE using RankT5-3B and the Gemma reranker.
While reranking generally improves effectiveness, reinforcing the motivation for distilling cross-encoders, we observe score drops on FEVER, Climate-FEVER, ArguAna, and SCIDOCS when reranking with RankT5.

This is interesting given rerankers’ typical advantages over retrievers, such as the ability to directly model query-passage relevance and RankT5-3B’s significantly larger size (3B parameters) compared to the BGE retriever (110M parameters).

However, the embedding models examined have undergone extensive fine-tuning. GTE and Arctic are trained on datasets like FEVER and other BEIR tasks, and include pretraining on scientific abstracts with corresponding titles, likely contributing to stronger effectiveness on datasets such as SCIDOCS. 
The training data for BGE is less clearly documented~\cite{bgecpack}, but is likely similarly aligned with BEIR evaluation tasks.
We further analyze model training data in Appendix~\ref{sec:studying_training_data}.

Overall, Table~\ref{tab:rerank_ndcg} highlights the potential of cross-encoder distillation and the challenges of fair evaluation, particularly when retrievers may have been pre-exposed to BEIR datasets, possibly to remain competitive on benchmarks like BEIR and MTEB.

\subsection{Passage De-duplication} 
Similar to previous work~\cite{pradeep2024ragnarokreusableragframework}, we identify many near-duplicate passages in MSMARCO and BEIR corpora, which may reduce generated query diversity and hinder contrastive learning.
We normalize text and remove passages that are substrings of others, eliminating many passages from the corpora studied, including over 950K from MSMARCO’s total 8.8M passages.

\section{Results and Analysis
}
\label{sec:eval_effectiveness}

\subsection{Listwise Distillation vs InfoNCE loss}

\begin{table*}[t]

    \centering
    \adjustbox{max width=0.7\textwidth}{
    \begin{tabular}{lccccc}
        \toprule
        \multirow{2}{*}{Query Type} & \multirow{2}{*}{\# Queries} & \multicolumn{2}{c}{DL19} & \multicolumn{2}{c}{DL20} \\
        \cmidrule(lr){3-4} \cmidrule(lr){5-6}
         & & NDCG & Recall & NDCG & Recall \\
        \midrule
        (No fine-tuning) & - & 56.3 & 52.6 & 54.6 & 62.1 \\
        \midrule
        User Queries (Human)             & 56K    & 71.6 & 63.9 & \textbf{73.4} & 73.0 \\
        User Queries Few-Shot (Synthetic)& 56K    & 71.7 & 64.9 & 70.3 & 72.1 \\
        User Queries Few-Shot (Synthetic)& 96K    & 72.8 & \textbf{65.3} & 72.8 & \textbf{74.5} \\
        User Queries Zero-shot (Synthetic)        & 56K    & 70.8 & 62.2 & 70.5 & 73.1 \\
        User Queries Zero-shot (Synthetic)        & 98K    & \textbf{74.0} & 62.9 & 71.8 & 73.7 \\
        \bottomrule
    \end{tabular}
    }
    \caption{Retrieval effectiveness for the E5-unsupervised model fine-tuned with human-written and synthetic queries. For the synthetic queries, results are provided for both a subset of 56K queries to provide a fair comparison and the full query set.}
    \label{tab:query_quality}
\end{table*}

Table~\ref{tab:loss_results} shows two notable findings.
First, training with a contrastive loss reduces effectiveness compared to the original model most of the time.
It is generally only E5-unsupervised that benefits from contrastive fine-tuning due to its lack of prior supervised fine-tuning.
Second, combining listwise distillation with a contrastive loss consistently outperforms using a contrastive loss alone across MSMARCO, SciFact, FiQA, TREC-COVID, and NFCorpus.
We focus on these datasets because, aside from MSMARCO, which serves as an in-domain benchmark, they represent datasets not seen during pretraining or fine-tuning of the base retrievers.
In most cases, the combined approach yields statistically significant improvements over using a contrastive loss alone.
The only case where contrastive learning outperforms using the combination of contrastive learning and listwise distillation is when evaluating the fine-tuned GTE model on TREC-COVID.
Given the advantages of the combined approach across models and datasets, we adopt it for all subsequent experiments.

\subsection{Corpus-Specific Fine-Tuning Effectiveness}

Table~\ref{tab:full_training} shows that both NDCG@10 and Recall@100 generally improve across all three models on most datasets.
However, there are notable exceptions, namely FEVER, Climate-FEVER, SCIDOCS, and ArguAna, where Section~\ref{sec:prior_tuning} highlighted challenges with the reranker's relative effectiveness.
These cases likely stem from prior fine-tuning of the retriever models on these datasets or related tasks, unlike the reranker.

There are also a few instances of marginal effectiveness drops after fine-tuning, such as DL19 NDCG for the GTE and Arctic models, FiQA recall for GTE, and DBPedia recall and TREC-COVID NDCG for Arctic.
While the combined listwise distillation and contrastive learning approach proves effective overall, these results present the difficulty of performing and evaluating embedding model fine-tuning.
To address these limitations, we explore using a stronger cross-encoder teacher with broader training coverage in Section~\ref{tab:full_general_training}.

The table also includes results from Promptagator, a more recent corpus-specific fine-tuning method for single-vector retrievers, scoring stronger than older work such as GPL~\cite{gpl}.
Notably, its effectiveness is generally below that of the base models even before fine-tuning, suggesting that recent models have outdone past methods even when fine-tuned on specific corpora.

\subsection{Evaluating Generated Queries}
\label{sec:eval_query_quality}

\subsubsection{Evaluating the Role of Query Types}

Table~\ref{tab:query_type} presents results from training models using different types of synthetic queries: zero-shot and few-shot natural user queries, questions, claims, titles, and keywords.
We again evaluate effectiveness across MSMARCO, FiQA, SciFact, TREC-COVID, and NFCorpus.

These datasets tend to include question-style queries.
For instance, MSMARCO queries are real-world search queries submitted to Bing, with non-questions filtered out during dataset construction~\cite{bajaj2016ms}.
FiQA queries are opinion-based and sourced from StackExchange, while TREC-COVID queries were written by people with biomedical training.
In contrast, SciFact queries are expert-authored scientific claims, and NFCorpus includes a mix of questions, titles, and descriptions from Q\&A forums, blogs, and videos.

Since some of the synthetic query types might have had more queries filtered, to ensure a fair comparison across query types when training, we downsample each set to match the smallest query pool, aligning on as many passages as possible.

Natural user search queries also tend to be diverse and not simply questions.
Interestingly, zero-shot natural user queries tend to yield the highest effectiveness scores, even in SciFact, where queries are scientific claims.
NFCorpus is the exception, where training with titles performs best, consistent with the dataset’s inclusion of title-like queries.

In general, training with most query types improves effectiveness, though there are a few exceptions: keyword queries reduce NDCG@10 on FiQA, and claim-based queries lower effectiveness on NFCorpus. On SciFact, training with questions, titles, keywords, or even the downsampled combination of all types slightly reduces NDCG@10.
However, it's worth noting that E5-unsupervised already performs exceptionally well on SciFact (NDCG@10 = 74.3), surpassing many leading BERT embedding models (see Table~\ref{tab:full_general_training}).

Training with all query types combined, without downsampling, consistently produces the strongest results across tasks.
This strategy likely benefits from training with more queries per passage, which may be particularly useful for smaller corpora like SciFact (5.2k passages) and NFCorpus (3.6k passages).
Moreover, this approach is query-type agnostic, making it more robust to variation in real-world or task-specific query distributions.

\begin{table*}[ht!]
\centering
\resizebox{\textwidth}{!}{
\begin{tabular}{l|cc|cc|cc|cc|cc|cc|cc}
\toprule
\textbf{Dataset} & \multicolumn{2}{c|}{\textbf{E5-base}} & \multicolumn{2}{c|}{\textbf{E5-base}} & \multicolumn{2}{c|}{\textbf{BGE-base}} & \multicolumn{2}{c|}{\textbf{GTE-base}} & \multicolumn{2}{c|}{\textbf{Arctic-m}} & \multicolumn{2}{c|}{\textbf{Distill}} & \multicolumn{2}{c}{\textbf{Distill}} \\
 & \multicolumn{2}{c|}{\textbf{unsupervised}} & \multicolumn{2}{c|}{\textbf{supervised}} &  &  & & & & & \multicolumn{2}{c|}{\textbf{RT5}} & \multicolumn{2}{c}{\textbf{(RT5, Gemma)}} \\

 & nDCG & Recall & nDCG & Recall & nDCG & Recall & nDCG & Recall & nDCG & Recall & nDCG & Recall & nDCG & Recall \\
\midrule
TREC-DL19          & 56.3 & 52.6 & 73.6 & 65.0 & 70.2 & 60.9 & 71.9 & 62.1 & 74.4 & 64.7 & 72.9 & 65.1 & 72.0 & 64.7 \\
TREC-DL20          & 54.6 & 62.1 & 73.2 & 73.4 & 67.7 & 71.5 & 71.5 & 69.8 & 72.1 & 74.2 & 71.5 & 74.9 & 74.9 & 76.0 \\
\midrule
TREC-COVID    & 60.6 & 11.9 & 76.5 & 13.2 & 78.1 & 14.1 & 75.3 & 14.0 & 82.2 & 14.8 & 78.3 & 13.9 & 79.0 & 14.2 \\
BioASQ        & 29.8 & 55.2 & 35.2 & 57.4 & 41.5 & 63.2 & 37.7 & 59.2 & 42.7 & 66.3 & 41.7 & 65.8 & 41.6 & 65.4 \\
NFCorpus      & 34.6 & 32.3 & 35.8 & 32.8 & 37.3 & 33.7 & 35.3 & 33.1 & 36.1 & 32.4 & 36.2 & 32.9 & 37.6 & 33.3 \\
NQ            & 37.6 & 80.1 & 51.7 & 87.2 & 53.1 & 94.2 & 46.2 & 84.4 & 53.9 & 90.0 & 49.4 & 86.8 & 52.1 & 87.9 \\
HotpotQA      & 52.4 & 68.3 & 60.7 & 75.4 & 72.6 & 87.3 & 62.9 & 74.7 & 70.7 & 87.5 & 66.6 & 81.7 & 73.1 & 86.8 \\
FiQA          & 40.1 & 71.8 & 36.3 & 67.0 & 40.6 & 74.2 & 48.7 & 81.7 & 42.5 & 74.8 & 39.8 & 71.2 & 40.9 & 73.4 \\
Signal-1M     & 24.1 & 29.2 & 24.6 & 25.5 & 28.9 & 31.1 & 25.9 & 28.3 & 30.4 & 33.2 & 28.7 & 29.5 & 29.0 & 31.3 \\
TREC-NEWS     & 44.6 & 45.8 & 39.0 & 42.5 & 44.2 & 49.9 & 40.2 & 43.1 & 48.3 & 50.3 & 44.1 & 47.1 & 45.7 & 48.2 \\
Robust04      & 41.2 & 32.7 & 38.4 & 29.4 & 44.4 & 35.1 & 47.7 & 35.7 & 45.0 & 34.0 & 50.1 & 38.0 & 50.1 & 39.6 \\
Arguana       & 38.7 & 89.5 & 50.6 & 96.7 & 63.6 & 99.2 & 62.1 & 99.2 & 56.5 & 98.4 & 53.8 & 97.4 & 55.1 & 97.9 \\
Touché-2020   & 21.8 & 37.8 & 29.4 & 41.6 & 25.7 & 48.7 & 27.5 & 48.3 & 33.2 & 50.0 & 33.8 & 49.0 & 34.1 & 48.9 \\
CQADupStack   & 34.4 & 67.7 & 34.6 & 64.0 & 42.4 & 76.2 & 36.1 & 72.0 & 42.0 & 74.8 & 35.8 & 67.8 & 36.1 & 68.2 \\
Quora         & 81.0 & 98.9 & 84.2 & 99.1 & 88.9 & 99.7 & 88.4 & 99.7 & 86.3 & 98.7 & 87.1 & 99.2 & 87.8 & 99.3 \\
DBPedia       & 34.5 & 46.1 & 39.1 & 49.7 & 40.7 & 53.0 & 36.9 & 46.9 & 44.7 & 58.7 & 43.2 & 56.4 & 44.6 & 58.2 \\
SCIDOCS       & 20.4 & 46.6 & 17.6 & 40.0 & 21.7 & 49.6 & 21.7 & 50.1 & 20.0 & 42.3 & 17.1 & 40.7 & 17.9 & 42.3 \\
FEVER         & 61.3 & 90.5 & 46.2 & 79.4 & 86.3 & 97.2 & 92.1 & 97.5 & 85.6 & 97.6 & 76.7 & 95.5 & 87.8 & 97.3 \\
Climate-FEVER & 15.6 & 41.4 & 11.5 & 38.4 & 31.2 & 63.6 & 40.1 & 71.7 & 34.7 & 66.7 & 24.1 & 59.4 & 37.2 & 67.3 \\
SciFact       & 74.3 & 98.7 & 71.2 & 96.9 & 74.1 & 96.7 & 75.5 & 97.3 & 70.5 & 94.8 & 72.4 & 96.7 & 74.2 & 97.3 \\
\midrule
\textbf{TREC-DL Average} & 55.5 & 57.4 & 73.4 & 69.2 & 69.0 & 66.2 & 71.7 & 66.0 & 73.3 & 69.5 & 72.2 & 70.0 & \textbf{73.5} & \textbf{70.4} \\
\textbf{BEIR (18) Average}    & 41.5 & 58.0 & 43.5 & 57.6 & 50.9 & \textbf{64.8} & 50.0 & 63.2 & \textbf{51.4} & 64.7 & 48.8 & 62.7 & 51.3 & 64.3 \\
\textbf{BEIR (13*) Average}    & 37.0 & 49.2 & 38.9 & 48.6 & 44.0 & 54.8 & 44.2 & 54.5 & 45.1 & 55.1 & 43.3 & 53.7 & \textbf{45.2} & \textbf{55.2} \\
\bottomrule
\end{tabular}
}
\caption{\textbf{nDCG@10} and \textbf{Recall@100} on TREC-DL and BEIR datasets. Distill-RT5 uses only the RankT5 cross‐encoder as a teacher, Distill-(RT5, Gemma) uses both RankT5 and the Gemma reranker as teachers. To calculate the BEIR (13*) average, we remove the NQ, HotpotQA, CQADupStack, FEVER, and Quora datasets due to the BGE, GTE, and Arctic models training on all or some of these datasets.}
\label{tab:full_general_training}
\end{table*}

\subsubsection{Human-written vs Synthetic Queries}

Table~\ref{tab:query_quality} compares the effectiveness of training E5-unsupervised using either human-written queries from MSMARCO or synthetic queries generated in either few-shot (given MSMARCO examples) or zero-shot settings.

Applying the filtering process from Section~\ref{sec:filtering}, human-written queries are filtered at a much higher rate, yielding only 56K usable queries compared to 96K (few-shot) and 98K (zero-shot) synthetic queries.
This is odd given that RankT5 was fine-tuned on these same human-written queries, suggesting that labeled MSMARCO passages are often not ideal matches for their queries, as noted by past work~\cite{rocketqa,ArabzadehSparseLabels}.

To ensure a fair comparison, we also train on a 56K subset of the synthetic queries, aligned as closely as possible to the same set of passages used in the human-written set. In this controlled setting, few-shot synthetic queries outperform the human-written ones on DL19, while zero-shot queries achieve slightly higher recall on DL20.
When training with the full synthetic sets (without downsampling), scores improve further. 
The only case where training with the human-written queries scores highest is DL20 NDCG\@10, where human-written queries retain the highest score.

Overall, these findings suggest no clear superiority between synthetic and human-written queries. Lightweight LLMs like Llama-3.1 (8B) can generate synthetic queries that are highly competitive for training, offering a practical and scalable alternative to collecting training queries.

\subsection{Training a General Retrieval Model}

Leveraging our findings on the effectiveness of combining listwise distillation with contrastive learning and utilizing diverse synthetic queries, we fine-tune an embedding model intended to compete with state-of-the-art BERT-base retrievers.

We fine-tune E5-unsupervised, which benefits from contrastive pre-training but lacks the supervised fine-tuning for retrieval present in the other models examined and the supervised E5 variant.

We train the model solely using synthetic queries and passages from MSMARCO, DBPedia, and FEVER, three corpora representative of broad, general-purpose domains: MSMARCO from Bing web search, FEVER from Wikipedia, and DBPedia from DBPedia’s structured entries.
This choice minimizes the risk of overfitting to narrow or specialized domains.

To obtain passages for listwise distillation and potential hard negatives, we aggregated the top-20 passages retrieved for each query by each of the baseline retrievers (BGE, GTE, and Arctic).
This pooled set was then reranked using the cross-encoder teachers, and the top 20 passages post-reranking were used for training.
Notably, our fine-tuning process uses only synthetic queries and teacher-derived relevance signals, contrasting with the baseline models (BGE, GTE, Arctic), which incorporate human-written queries and relevance labels from various datasets in their training.

We explore two distillation setups: (1) Distill-RT5 using RankT5-3B alone as the teacher, and (2) Distill-(RT5, Gemma) combining RankT5-3B with the Gemma-reranker, using a weighted sum of their scores ($0.25 \times$ RankT5 score $+$ $0.75 \times$ Gemma score).
We do not closely tune this ratio, but find that it is helpful to combine the scores from both for distillation.
All models are trained using the same hyperparameters described in Section~\ref{sec:hyperparameters}.
Table~\ref{tab:full_general_training} summarizes the results.

Distill-RT5 shows strong effectiveness.
Compared to both the unsupervised and supervised E5 variants, Distill-RT5 scores much stronger on BEIR on average.
However, reflecting the limitations observed in Table~\ref{tab:rerank_ndcg}, it underperforms on datasets where RankT5 struggles, such as FEVER, Climate-FEVER, SCIDOCS, and ArguAna.

Distill-(RT5, Gemma), which we release as \textit{cadet-embed-base-v1} delivers strong effectiveness across both in-domain and out-of-domain settings.
On the TREC-DL benchmark, which serves as a fair in-domain evaluation for all models, it achieves the highest average scores, outperforming all baselines.
On BEIR, Distill-(RT5, Gemma) remains highly competitive, achieving 51.3 NDCG@10 (shy of Arctic’s 51.4) and 64.3 Recall@100 (compared to BGE’s 64.8). 
However, the baseline models considered are trained on subsets of BEIR, including NQ, HotpotQA, CQADupStack, FEVER, and Quora, introducing potential bias (see Appendix~\ref{sec:studying_training_data}).
When these datasets are excluded, our model achieves the highest average scores across the remaining BEIR tasks.

\label{sec:general}

\section{Concluding Discussion}

Our work shows that fine-tuning with conventional contrastive learning alone can often worsen the retrieval effectiveness of state-of-the-art embedding models on new corpora or tasks.
However, we find that incorporating cross-encoder listwise distillation is promising, delivering effectiveness gains across varied datasets and models.
We argue that this listwise distillation approach has been under-explored in recent work, and by leveraging it, we trained a BERT-base embedding model, achieving state-of-the-art results among comparable models.

Furthermore, we show that synthetic queries can deliver comparable utility to human-written queries for training purposes.
Generating a diverse set of synthetic query types proves to be an effective strategy for adapting retrievers to specific corpora, yielding benefits while being agnostic to the query types used in evaluation datasets.

We successfully trained highly effective embedding models using a purely synthetic data pipeline, from generated queries to relevance signals provided by teacher cross-encoders, offering a practical and scalable path for improving dense retrieval.

\newpage

\section*{Limitations}

While our focus on BERT-base embedding models facilitated efficient training and direct comparison with strong baselines, applying our methods to larger models such as LLM-based backbones or to newer encoder models like ModernBERT~\cite{warner2024smarter} represents a logical next step for potentially advancing retrieval effectiveness further and demonstrating the effectiveness of our approach.

We also limit our evaluation to three supervised embedding models and one unsupervised baseline (E5-unsupervised). 
In some experiments, we restrict the number of datasets to focus on specialized domains and avoid contamination from overlapping training data.
While it would be nice to evaluate our methods with even more datasets and more models, we do our best to present fair and extensive experiments to support our claims, given limited computational resources.

We also do not focus on multilingual retrieval.
However, Llama-3.1 (8B) is capable of generating multi-lingual queries, and the Gemma-reranker used in this work~\cite{limaking} is capable of multi-lingual reranking.
So our approach should allow for training multi-lingual retrieval models.

While our approach is effective, there are other possible techniques that we do not explore.
Distilling from highly effective listwise rerankers~\cite{pradeep2023rankzephyr, tamber2023pre, zhang2023rankwithoutgpt} is one promising direction, although these rerankers output rankings rather than the continuous scores useful for our current distillation method.

\bibliography{custom}
\bibliographystyle{acl_natbib}

\newpage

\appendix

\label{sec:appendix}

\section{Analyzing the Training Data of the Retrieval Models Studied}
\label{sec:studying_training_data}
The complex and varied pre-training and fine-tuning data used for the BGE, GTE, and Arctic baseline models makes tracking precise dataset or task-specific contributions difficult.
In many cases, it is unclear what particular data is used for training these models.
However, there seems to be a heavy overlap in the training data of the models and subsets for BEIR as a possible result of the developers of these models aiming to have their model appear competitive, risking not only task contamination but also test data contamination.

GTE~\cite{gte} trains using BEIR datasets such as MSMARCO, NQ, HotpotQA, FEVER, and Quora as well as various datasets released in MEDI~\cite{su-etal-2023-one} and BERRI~\cite{asai-etal-2023-task}, though it is not clear what data in particular is used.
The training data mix for GTE also includes StackExchange data for duplicate question retrieval, which matches the evaluation used in the CQADupStack subset for BEIR.
The pre-training data, including academic paper abstracts, Twitter posts, and news articles, also seems to overlap heavily with BEIR evaluation. 

Snowflake's Arctic embedding model~\cite{merrick2024arctic} trains using NQ, HotpotQA, FEVER, and title-body pairs from 
StackExchange.
The model is also trained on title-body pairs from scientific papers.

At the time of writing, the full extent of the BGE embedding model's training data is not clear, and a link leading to details about the model's training seems to be dead\footnote{\url{https://data.baai.ac.cn/details/BAAI-MTP}}.
The documentation of MTEB\footnote{\url{https://github.com/embeddings-benchmark/mteb/blob/main/mteb/models/bge_models.py}} suggests that the training data of the BGE model examined includes MSMARCO, NQ, HotpotQA, and Quora, as well as title-body pairs from 
StackExchange.

In Table~\ref{tab:full_general_training}, we specifically drop MSMARCO, NQ, HotpotQA, FEVER, and CQADupStack in the calculation of the BEIR(13*) average for a fair comparison. However, we do note that other datasets not excluded present challenges as well, such as Climate-FEVER due to its similarity with FEVER and SCIDOCS due to models training on title-body pairs from scientific papers. 

We also acknowledge that the Gemma reranker, which we distill from into our Distill-(RT5, Gemma) model, is trained on a broad mix of datasets, including MSMARCO, NQ,  ArguAna, HotpotQA, FEVER, as well as other datasets such as PubMedQA.
However, we do our best to limit task contamination to our Distill-(RT5, Gemma) model by training it only on general web corpora and only with our synthetic queries.
In general, we observe that our model typically underperforms the BGE, GTE, and Arctic models on these particular datasets.

\section{Tuning the Threshold for Filtering Hard-Negatives for Contrastive Learning}
\begin{figure}[ht]
    \centering
    \includegraphics[width=0.95\columnwidth]{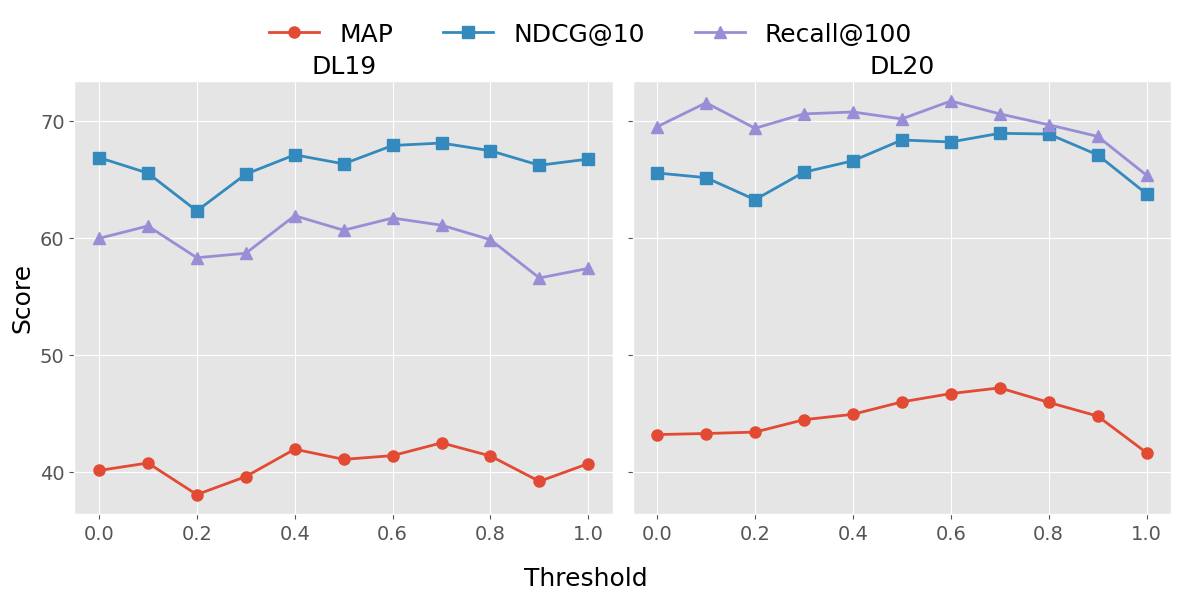} 
    \caption{Retrieval effectiveness scores on DL19 and DL20 at different hard negative filtering thresholds during E5-unsupervised fine-tuning on MSMARCO passages and synthetic queries.}
    \label{fig:threshold_tuning}
\end{figure}

Similar to past work,~\cite{nvretriever}, we tune the threshold for filtering hard-negatives.
Instead of using scores from a teacher embedding model, we use the cross-encoder’s normalized scores.
To tune the threshold, we train the E5-unsupervised model on MSMARCO passages and the corresponding generated queries.
We evaluate the models trained with different thresholds on DL19 and DL20.
We find that 60\% is a reasonable threshold that balances the retrieval metrics well.
If a hard-negative passage has a score above 60\% of the score for the positive passage that was used to generate the synthetic query, we do not use the passage as a hard-negative in the contrastive loss.
Figure~\ref{fig:threshold_tuning} shows weaker effectiveness scores at either extreme for thresholding, and a rough peak around 0.6.

\section{Computational Costs}

All experiments were conducted using BERT-base models (110M parameters or 137M in the case of gte-base-en-v1.5 because of its slight modifications) and on single 48GB GPUs (NVIDIA RTX 6000 Ada or L40S, depending on availability).
Key computational steps included query generation, reranking, and model training.
Using vLLM~\cite{kwon2023efficient} for inference on an RTX 6000 Ada, generating 100k synthetic queries for MSMARCO passages took approximately 1-2 hours.
Subsequently, reranking the top 20 passages for these queries required roughly 5 hours on the same GPU.
Model training times varied by dataset, with the MSMARCO fine-tuning task taking 48-60 hours per model.
Note that times can vary based on passage lengths.

\end{document}